\documentclass[twocolumn,showpacs,preprintnumbers,amsmath,amssymb, superscriptaddress]{revtex4}

\usepackage{graphicx}
\usepackage{dcolumn}
\usepackage{bm}
\usepackage{color}

\begin{document}


\title{
Quantification of group chasing and escaping process
}

\author{Shigenori Matsumoto}
 \affiliation{Department of Applied Physics, Graduate School of Engineering,
 The University of Tokyo, 7-3-1, Hongo, Bunkyo-ku, Tokyo 113-8656, Japan.}
  \author{Atsushi Kamimura}
 \affiliation{Institute of Industrial Science, 
 The University of Tokyo, 4-6-1, Komaba, Meguro-ku, Tokyo 153-8505, Japan.}
 \author{Tomoaki Nogawa}
 \affiliation{Department of Applied Physics, Graduate School of Engineering,
 The University of Tokyo, 7-3-1, Hongo, Bunkyo-ku, Tokyo 113-8656, Japan.}
 \author{Takashi Shimada}
 \affiliation{Department of Applied Physics, Graduate School of Engineering,
 The University of Tokyo, 7-3-1, Hongo, Bunkyo-ku, Tokyo 113-8656, Japan.}
  \author{Nobuyasu Ito}
 \affiliation{Department of Applied Physics, Graduate School of Engineering,
 The University of Tokyo, 7-3-1, Hongo, Bunkyo-ku, Tokyo 113-8656, Japan.}
 \author{Toru Ohira}
 \affiliation{Sony Computer Science Laboratories, Inc., 3-14-13, Higashi-gotanda, Shinagawa-ku,  Tokyo, 141-0022, Japan.}
\date{\today}
             
\begin{abstract}

We study a simple group chase and escape model by introducing new parameters with which configurations of chasing and escaping in groups are classified into three characteristic patterns.
In particular, the parameters distinguish two essential configurations: a one-directional formation of chasers and escapees, and an escapee surrounded by chasers.
In addition, pincer movements and aggregating processes of chasers and escapees are also quantified.  
Appearance of these configurations highlights efficiency of hunting during chasing and escaping.

\end{abstract}


\maketitle

Collective phenomena, such as school of fish and formation flight of birds, have been attracting interests for a long time.
Recently, problems on various self-propelled particles have been extensively researched in the context of physics\cite{vicsek_review, vicsek, romanczuk}.
Previous works mainly focused on spontaneous group-formation in experiments and mathematical models.
Indeed, group motions often become crucial for survival in nature.
For example, collective motions in hunting have advantages to chase a prey for a predator as well as to escape from a predator for a prey.
Hunting is a widespread phenomenon observed from micro-organisms such as a neutrophil capturing viruses\cite{neutrophil} to macroscopic entities in animal societies\cite{hunting} and robotics\cite{robot1,robot2}.
Preys and predators adopt various patterns of collective motions as a result of competition for survival.
A way to select an optimal strategy for each group is an interesting point.
In a hunting motion, basic actions are chases and escapes.
A problem of one-on-one chase-and-escape also has attracted mathematicians with its complex trajectories even in simple setups\cite{isaacs65,nahin07}.
However, group hunting could not be explained just by summing up of individual performances of prey or predator and can be quite more complex.
For example, cooperation within predator or prey group can give more diverse strategies for chasing or escaping.
A pincer movement is one typical example.
The emergence of collective motions in group hunting has also been attracting interests, however quantitative study has not been done.

\begin{figure}[t]
    \centering
    \includegraphics[width=\linewidth]{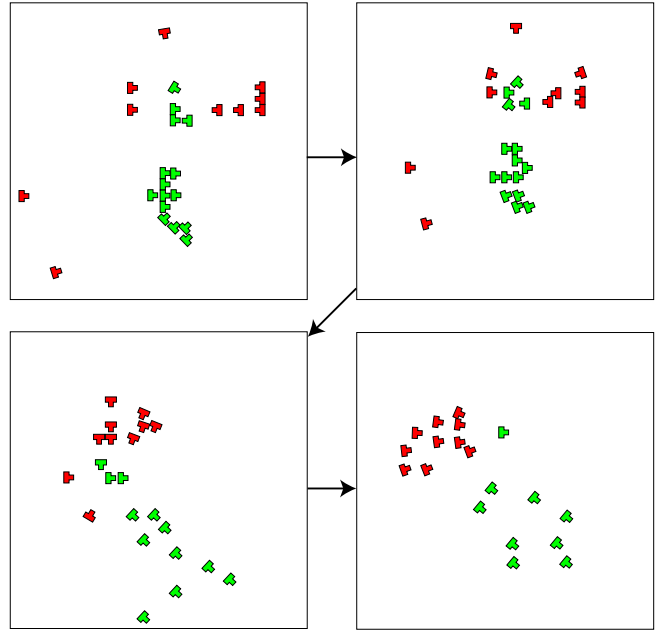}
    \caption{(color online)Capturing and escaping pattern in group. Red and green convex objects are chasers and escapees, respectively. The convex direction corresponds to a traveling direction.}
    \label{snapshots}
\end{figure}

One example of capturing and escaping patterns is shown in Fig.~\ref{snapshots} which is actually a result of numerical simulation described later.
Chasers surround aggregate of escapees to capture them.
On the other hand, a mass of escapees gets away from chasers while they chase ``decoy" escapees.
These patterns of group motions seem universal behaviors in hunting.
If these collective motions have an advantage rather than one-on-one hunting,
animals may adopt those as simple strategies through natural selection process.
To investigate the emergence of hunting patterns, evaluation of group hunting is important viewpoint.

\begin{figure}[t]
    \centering
    \includegraphics[width=\linewidth]{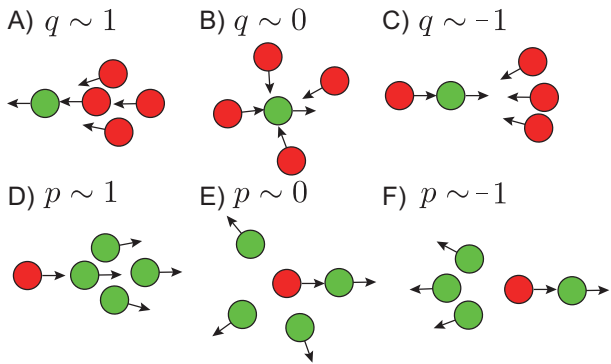}
    \caption{Chasing and escaping processes characterized by parameters $q$ and $p$. The pattern A, B, and C represent motion of chasers around an escapee with corresponding values of $q$. 
    In contrast,  The pattern D, E, and F represent motion of escapees around a chaser with corresponding values of $p$. }
    \label{chase_and_escape}
\end{figure}

First, we consider simple and feasible patterns in order to characterize chases and escapes.
We intuitively classify collective motions of capturing into several patterns.
As a simple classification, we consider capturing patterns by focusing on a single escapee or a single chaser with
its relation to opponents.
The first row of Fig.~\ref{chase_and_escape} shows chaser's motions while many chasers pursue one escapee (A to C), 
and the second row shows escapee's motions while many escapees escape from one chaser (D to F).
A situation of each pattern is explained as follows:
A) A number of chasers follow an escapee in a one-directional formation.
 B) Chasers surround escapees.
Capturing of the escapees typically follow this pattern.
C) One chaser drives an escapee into a group of chasers. This is a transient behavior often observed leading to the 
pattern B.
 D) Escapees escape in a one-directional formation from a single chaser.
 E) Escapees scatter away from a chaser isotropically or are divided into small groups. 
 This pattern frequently appears right after a chaser invades into an aggregate of escapees.
 F) One chaser runs after an escapee, while nearby escapees escape in different directions of the chaser.

\begin{figure}[t]
    \centering
    \includegraphics[width=\linewidth]{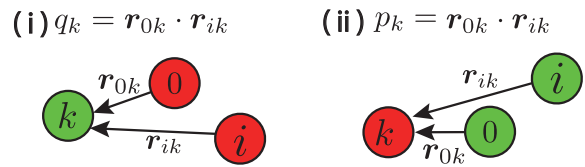}
    \caption{Illustrations for calculating parameters (i) $q_k$ and (ii) $p_k$ in case of three players. }
    \label{definition}
\end{figure}

Let us introduce a parameter $q$ to distinguish the pattern A to C, which is assigned to escapees. 
At each time step, we focus on every escapee $k$, and the $n^C_{k} + 1$ chasers chasing the escapee $k$. 
Here, we index the nearest chaser as $i=0$ and other chasers as $i=1, ..., n^C_k$.
For each escapee $k$, we define the parameter $q_k$ as 
\begin{equation}
q_k =\frac{1}{n^C_{k}}\sum^{n^C_k}_{i=1} \hat{r}_{ik} \cdot \hat{r}_{0k}, 
\end{equation}
where $\hat{r}_{0k}$ denotes a unit vector pointing the direction from the nearest 
chaser $i=0$ to the escapee $k$, while $\hat{r}_{ik}$ are also unit vectors 
from the $i$-th chaser to the escapee $k$.
Figure \ref{definition}(i) illustrates $q_k$ when the escapee $k$ is chased by two chasers.
By this parameter, the three patterns A, B, and C yield $q_k \sim 1, 0$ and $-1$, respectively.
We also introduce the average of $q_k$ for $\tilde{N}_T$ escapees which are chased by more than two chasers as
\begin{equation}
\bar{q} = \frac{1}{\tilde{N}_T}\sum^{\tilde{N}_T}_{k=1}q_k.
\end{equation} 

We also distinguish the patterns D to F from the viewpoint of chasers rather than escapees by the similar expression to $q$.
At each time step, we focus on each chaser $k$, and escapees escaping from the chaser. 
A parameter for the chaser $k$ is defined in the same way as, 
\begin{align*}
p_k&=\frac{1}{n^T_k }\sum^{n^T_k}_{i=1} \hat{r}_{ik} \cdot \hat{r}_{0k} \\
\bar{p} &= \frac{1}{N_C}\sum^{N_C}_{k=1}p_k,
\end{align*}
where the $n^T_k$ denotes the number of escapees escaping from the chaser $k$.
Figure \ref{definition}(ii) also illustrates $p_k$ when two escapees are escaping from the chaser $k$.
By this parameter, the three patterns D, E, and F in Fig.~\ref{chase_and_escape} yield $p_k \sim 1, 0$ and $-1$, respectively.

\begin{figure}[t]
    \centering
    \includegraphics[width=\linewidth]{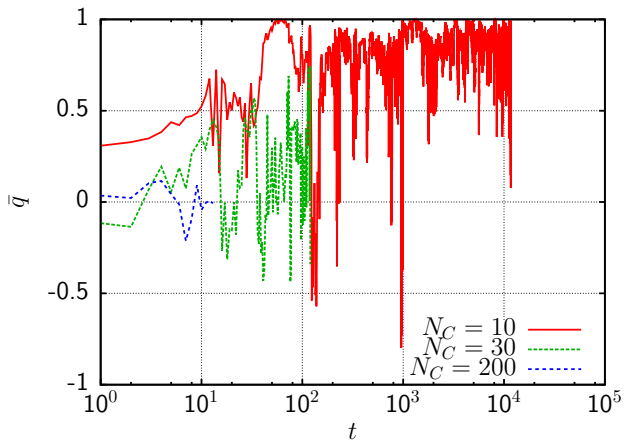}
    \caption{Time evolution of parameter $\bar{q}$ for the numbers of chaser, $N_C=10,30,200$.  Initially 10 targets are randomly placed on the $100\times100$ square lattice.}
    \label{time_q}
\end{figure}

The above parameters $q$ and $p$ quantitatively distinguish the patterns shown in Fig.~\ref{chase_and_escape}.
By these parameters, we can quantify  group formations locally from snapshots such as Fig.~\ref{snapshots}.
From another point of view, this analysis corresponds to local structure with anisotropy.
Therefore, we consider that the parameter is also useful to characterize the anisotropic structure and 
can be applied to a variety of physical systems, such as amorphous materials\cite{torquato}.

Next, we demonstrate an application of the present parameter to simple model of group hunting.
Recently, a simple model of chase and escape in groups is proposed\cite{kamimura10}.
Even this simple model shows intriguing motions related to collective hunting behavior. 
Figure \ref{snapshots} actually shows snapshots of a simulation of the model.
In the previous work, macroscopic quantities, the time for entire catch $T$ and typical lifetime $\tau$ of escapees,
 are observed but the dynamics is not understood in relation with the chasing spatial configuration in a group.
We investigate details of collective motions for chase and escape in a group by $q$ and $p$.

Let us briefly explain the model in Ref.~\cite{kamimura10}. 
Initially, two types of players, named chasers and targets (escapees), are placed randomly over the two-dimensional square lattice. A periodic boundary condition is imposed on the lattice.
A chaser moves by one lattice unit toward its nearest target.
On the other hand, a target tries to move away from its nearest chaser by one lattice unit.
Here, the nearest player means the one located in the shortest Euclidean distance, 
and if there are multiple nearest chasers (targets), the target (chaser) chooses one of them with equal probabilities.
Each player chooses the next hopping site in the following ways.
When a chaser and its nearest target are on the same axis, the chaser chooses the nearest site toward the opponent,
 but the target randomly chooses one of three neighboring sites to increase the distance.
In the other situation, chasers and targets choose one of two possible nearest sites with an equal probability in order to move closer to,
 or away from their opponents, respectively.
 We include exclusion volume effect such that players remain in the same site if the next hopping site is occupied. 
When a chaser and a target are placed next to each other, then the chaser moves to the position of the target to remove it. 
This catching rule leads to monotonic decrease of the number of targets.
After the catch, the chaser pursues one of the remaining targets in the same manner.
The simulation starts with $N^0_T$ targets and $N_C$ chasers, and end when all the targets are caught.

Figure \ref{time_q} shows time evolution of $\bar{q}$ for different numbers of chasers with $N_T^0=10$.
When the number of chasers is much larger than initial number of targets ($N_C=200$), 
almost all targets are initially surrounded by chasers.
Thus the initial $\bar{q}$ is close to 0.
Since most of targets are immediately caught in few time steps by the pattern B, $\bar{q}$ remains almost cnstant.
On the other hand, when the number of chaser is as small as the initial number of target ($N_C=10$),
$\bar{q}$ initially fluctuates, and eventually approaches 1.
This result indicates that the remaining targets are generally chased by a group of chasers as in the pattern A.
This can be explained as follows.
The catching event results in aggregation of chasers as in the pattern B.
After catch, the gathering chasers tend to pursue the same nearest-target, leading to the pattern A, as the remaining targets become small.
For an intermediate number of chaser ($N_C=30$), behavior of $\bar{q}$ depends on an initial configuration.
In particular, the initial position of players influences appearance of the pattern A.
A fraction of samples in which the pattern A appears increases as the number of chaser decreases.

\begin{figure}[b]
    \centering
    \includegraphics[width=\linewidth]{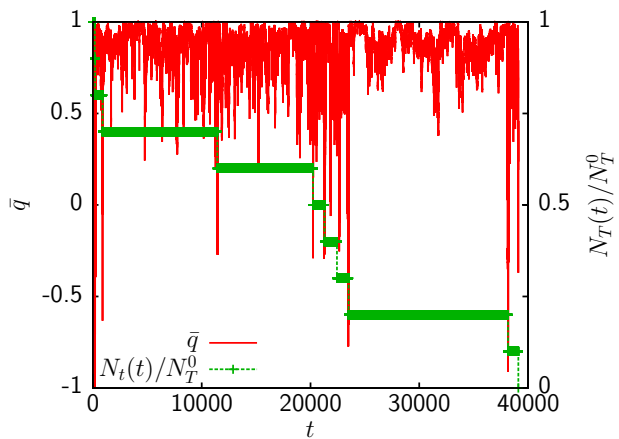}
    \caption{Time evolution of $\bar{q}$ and $N_T(t)/N_T^0$ with $N_T^0=10$ and $N_C=5$.}
    \label{decay_q2}
\end{figure}

Now let us turn our attention to the relation between time evolution of $\bar{q}$ and the number of target, $N_T(t)$.
Figure \ref{decay_q2} shows the time evolution of $\bar{q}$ and $N_T(t)$ for $N_T^0=10$ and $N_C=5$.
We find that $\bar{q}$ fluctuates around the value 1 in most of time, but spike-like dips are observed right before catch events which correspond to the decrement timings of targets.
This indicates that chasers surround targets to capture them in the patterns B and C.
At the capturing motion, the chasers aggregate, and after that they form a larger group chasing a single nearest target.
This leads to the rapid rise of $\bar{q}$ to 1.

\begin{figure}[t]
    \centering
    \includegraphics[width=\linewidth]{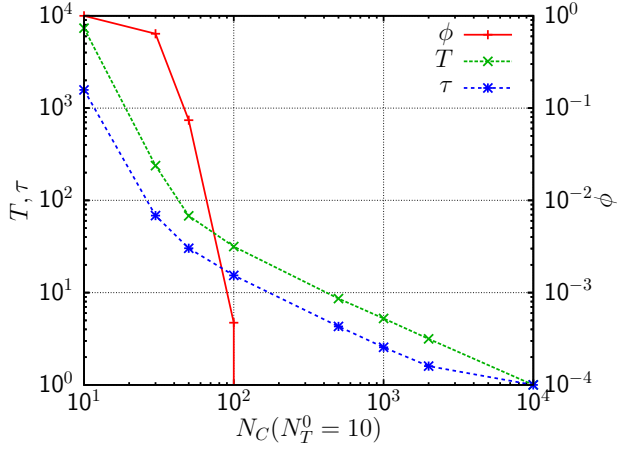}
    \caption{Dependence of $T$, $\tau$ and $\phi$ on $N_C$. $\phi$ denotes the ratio of samples which achieve $\bar{q}>0.8$.   }
    \label{q_T_tau}
\end{figure}

This parameter $\bar{q}$ also explains the crossover behavior of $T$ and $\tau$ as a function $N_C$.
Here, $T$ is defined as a time in which all targets are caught and $\tau$ is given as $\tau = \sum_{t=0}^{T} t(N_T(t-1) - N_T(t))/N_T^0$.
As shown in Fig. \ref{q_T_tau}, $T$ and $\tau$ show two kinds of power-law behaviors as a function of $N_C$ and kinks around $N_C=50$ for $N_T^0 = 10$.
With parameter $q$, we can understand now that these crossovers come from difference of frequency in appearance of the patterns A, B and C.
As explained before in Fig.~\ref{time_q}, $\bar{q}$ stays near 0 (the pattern B) when $N_C$ is much larger than $N_T^0$,
and $\bar{q}\sim 1$ (the pattern A) appears frequently when $N_C$ becomes smaller.
Here, we quantify them by observing $\phi$: the fraction of samples which achieve $\bar{q} > 0.8$ at least once until capturing all targets.
In Fig.~\ref{q_T_tau}, 
we clearly see that the crossover point coincides with the point $\phi$ becomes almost zero.
In other words, the pattern A dominates below this crossover point, while almost all targets are rapidly captured with the pattern B or C after this point.

\begin{figure}[t]
    \centering
    \includegraphics[width=\linewidth]{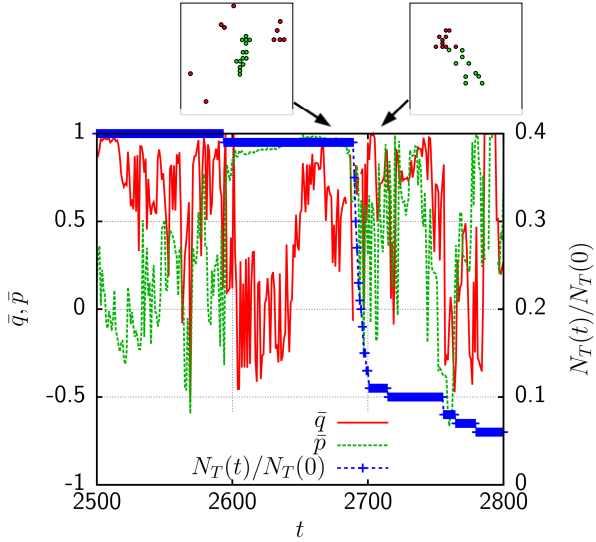}
    \caption{Time evolution of $\bar{q}, \bar{p}$ and $N_T(t)/N_t^0$ in the vicinity of a drastic decay of the numb of the targets with $N_T^0=50$ and $N_C=10$.   }
    \label{decay_q_p}
\end{figure}

The parameter $q$ quantifies how single target is pursued by a group of chasers. 
However, it is insufficient to quantify behaviors of aggregating targets.
In particular, with the condition $N_T^0 > N_C$, a few drastic decreases in $N_T$ are observed as shown later.
This suggests that an aggregate of targets is caught at once surrounded by chasers.
In such situation, the parameter $p$, together with $q$, helps us to understand the drastic decrease of the number of the targets.
As an example, Fig.~\ref{decay_q_p} shows time evolution of $\bar{p}, \bar{q}$ and $N_T(t)$ for $N_T^0=50$ and $N_C=10$.
The drastic decrease of $N_T(t)$ occurs around $t=2700$.
Before the catching event, $\bar{p}$ keeps around 2 for a certain period.
This indicates that aggregate of targets is caught by surrounding chasers.
The upper figures in Fig.~\ref{decay_q_p} show snapshots before and after the event.
After the drastic decay, $\bar{p}$ rapidly decrease to 1 or less exhibiting the pattern E or F.
Simultaneous occurrences of such events of $\bar{p}$ and drastic decay of $N_T(t)$ explains the collective catching motion by the surrounding targets.

\begin{figure}[t]
    \centering
    \includegraphics[width=\linewidth]{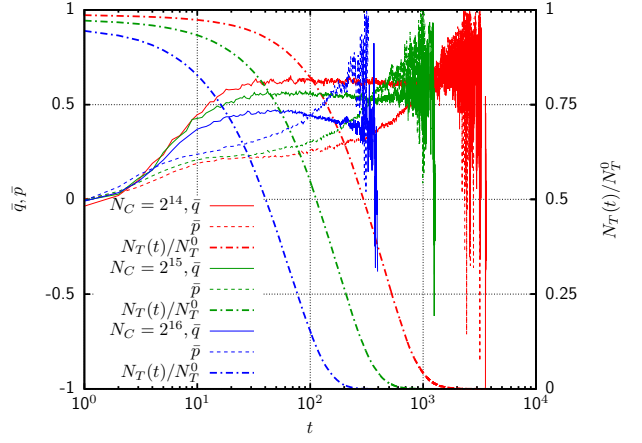}
    \caption{Time evolution of $\bar{q}, \bar{p}$ and $N_T(t)/N_T^0$ with $N_T^0=2^{18}, N_C=2^{14},2^{15},2^{16}$.  The system size is $2048 \times 2048$.  }
    \label{large_system}
\end{figure}

So far, we have shown results of rather small size system where $\bar{p}$ and $\bar{q}$ approximately represent the individual chasing and escaping processes.
In order to investigate the average (dominant) behavior of players in a given condition,
 we investigate trend of group motion in group chase and escape with a large amount of players.
For this purpose, we expand the system size keeping the number density of players and effectively take the ensemble averages of the parameters.
Here, the system size is set to $2048\times2048$ square lattice.
We note that individual motions of local aggregates are no longer captured by the parameters, however they reflect the entire spectrum.
As shown in Fig.~\ref{q_T_tau}, $\phi$ becomes zero for  $N_C/L^2 > 0.01$, 
where $L$ denotes a linear system size.
On the other hand, for higher density of targets, $\phi$ shows a finite value.
Figure \ref{large_system} shows time evolutions of $\bar{q}$ and $\bar{p}$ with $N_T^0=2^{18}, N_C=2^{14},2^{15}$ and $2^{16}$.
Even in the case of $N_T^0 > N_C $, the pattern B is initially dominant for the chased targets as $\bar{q} \sim 0$.
In addition, $\bar{p} \sim 0$ indicates that dominant escaping is in the pattern E, which means that most of targets can survive in a while.
In the middle stage, plateau region appears in $\bar{q}$ and $\bar{p}$ for all $N_C$.
In that region, $\bar{q}$ increase $N_C$ decreases.
In contrast, $\bar{p}$ is greater than 0 but far less than 1, and targets escape almost in the pattern E and occasionally with the pattern D.
This behavior of $\bar{q}$ and $\bar{p}$ tell us that, in the plateau region, a fair number of targets escape while chasers chase ``decoy" targets.

After the plateau region, $\bar{q}$ and $\bar{p}$ start to fluctuate as the remaining targets become small.
It is noteworthy that the value of $\bar{q}$ becomes closer to 1 even though $N_C/L^2 > 0.01$. 
This is not observed when the initial density of targets is low.
It indicates that chasers and targets are likely to take the pattern A because chasers chase ``decoy" targets at the initial and the plateau stages. 
The appearance of the pattern A contributes to relatively long lifetime of target as in Fig.~\ref{q_T_tau}.
This fact corresponds to our previous result\cite{grannada_matsumoto}, which showed the lifetime of targets is longer in a high-density case than that in a low-density case for $N_T^0$.

In conclusion, we investigate group behavior of chasing and escaping by two parameters we proposed, and 
classify the motions into three patterns from a target's and a chaser's viewpoints.
By increasing the number of chasers, the kinks of power-law behaviors for the entire catch $T$ and typical lifetime $\tau$ of targets appear because of the one-directional formation, the pattern A. 
On the other hand, collective behavior of targets is also characterized. 
By observing time evolutions of $q$ and $p$, we can detect the drastic catch-at-once of targets' aggregate by the surrounding chasers.
In addition, we find that the one-directional formation causes longer lifetime of targets.

The basic idea of the present parameters is to quantify spatial distribution of related objects from a view point of fine subject, 
and to characterize local structures with anisotropy.
Our  parameters are defined without any dynamical properties of objects.
Then we expect that they can be applied to various systems.
In particular, they have advantages to investigate structures of multi-component systems such as the chasing and escaping entities, 
in which a characteristic property appears relatively locally but not in a global scale.
A randomly packed media can be another example and analyzing the local structures could be promising by applying the parameters\cite{matsumoto}.

This work was partly supported by Award No. KUK-I1-005-04 made by King Abdullah University of Science and Technology (KAUST).


\begin{thebibliography}{999}
\bibitem{neutrophil} D. Rogers, Crawling Neutrophil Chasing a Bacterium (1950), http://www.biochemweb.org/neutrophil.shtml.
\bibitem{hunting} L. A. Dugatkin, {\it Cooperation among animals: an evolutionary perspective.}, Oxford University Press (1997).
\bibitem{robot1} J. P. Hespanha, H. J. Kim, and S. Sastry, Proc. 38th Conference on Decision and Control, p. 2432 (1999).
\bibitem{robot2} R. Vidal, O. Shakernia, J. H. Kim, D. H. Shim, S. Sastry, IEEE Trans. Robotics and Automation 18, 662 (2002).
\bibitem[Isaacs(1965)]{isaacs65} R.~Isaacs, \emph{John Wiley \& Sons, New York}  (1965).
\bibitem[Nahin(2007)]{nahin07}P.~J. Nahin, \emph{Princeton University Press, Princeton}  (2007).
\bibitem{vicsek_review} T. Vicsek and A. Zafiris, arXiv:1010.5017 (2010).
\bibitem{torquato} S. Torquato and F. Stillinger, Rev. Mod. Phys. 82, 2633-2672 (2010).
\bibitem{vicsek} T. Vicsek, A. Czirok, E. Ben-Jacob, I. Cohen, O. Shochet, Phys. Rev. Lett. 75, 1226 (1995).
\bibitem{romanczuk} P. Romanczuk, I. D. Couzin, and L. Schimansky-Geier, Phys. Rev. Lett. 102, 010602(2009).
\bibitem{kamimura10} A. Kamimura and T. Ohira, New J. Phys. 12, 053013 (2010).
\bibitem{grannada_matsumoto} S. Matsumoto, T. Nogawa, A. Kamimura, N. Ito, and T Ohira, AIP Conf. Proc. 1332, 226-227 (2011).
\bibitem{Bernal} J. D. Bernal, Proc. R. Soc. London, Ser. A  280, 299 (1964).
\bibitem{matsumoto} S. Matsumoto, T. Noagawa, T. Shimada, and N. Ito, arXiv:1005.4295 (2010).
\end{thebibliography}
\end{document}